\documentclass[aps,prl, twocolumn,footinbib,
letterpaper,showpacs]{revtex4}

\usepackage{times}
\usepackage{graphicx}
\usepackage{epsfig}
\usepackage{subfigure}
\usepackage{amssymb,amsmath}
\usepackage{hyperref}
\usepackage{setspace}
\usepackage{url}

\hypersetup{
    colorlinks = true,
    linkcolor = blue,
    anchorcolor = blue,
    citecolor = blue,
    filecolor = blue,
    pagecolor = blue,
    urlcolor = blue
}

\begin{document}

\vspace{4.0cm}

\title{Light  Stop from $b$-$\tau$ Yukawa Unification}

\author{Ilia Gogoladze }
\altaffiliation{On leave of absence from: Andronikashvili Institute of Physics, GAS, Tbilisi, Georgia.}
\affiliation{Bartol Research Institute, Department of Physics and Astronomy\\
University of Delaware, Newark, Delaware 19716}

\author{Shabbar Raza}
\altaffiliation{On study leave from: Department of Physics, FUUAST,  Islamabad, Pakistan}
\email[E-mail:]{shabbar@udel.edu}
\affiliation{Bartol Research Institute, Department of Physics and Astronomy\\
University of Delaware, Newark, Delaware 19716}

\author{Qaisar Shafi}
\affiliation{Bartol Research Institute, Department of Physics and Astronomy\\
University of Delaware, Newark, Delaware 19716}

\pacs{12.60.Jv, 12.10.Dm, 14.80.Ly}

\begin{abstract}
{\noindent}We show that $b$-$\tau$ Yukawa unification can be successfully implemented in the constrained minimal supersymmetric model and it yields the stop co-annihilation scenario. The lightest supersymmetric particle is a bino-like dark matter neutralino,
which is accompanied by a 10-20\% heavier stop of mass $\sim 100-330 \,{\rm GeV}$. We highlight some benchmark points which show
a gluino with mass $\sim 0.6 - 1.7 \, {\rm TeV}$, while the first two family squarks and all sleptons have masses in the multi-${\rm TeV}$ range.

\end{abstract}

\maketitle

The apparent unification at $M_{\rm G} \approx  2\times 10^{16} \,{\rm GeV}$ of the three Standard Model (SM) gauge couplings, assuming TeV scale
supersymmetry (SUSY), strongly suggests the existence of an underlying grand unified theory with a single coupling constant.
The minimal supersymmetric SU(5) model, in addition to unifying the gauge couplings, also predicts unification at
$M_{\rm G}$ of the third family bottom ($b$) quark and tau lepton ($\tau$) Yukawa couplings \cite{BottomTauNonSUSY}.
This $b$-$\tau$ Yukawa unification (YU) is to be contrasted with the minimal supersymmetric $SO(10)$ and  $SU(4)_c\times SU(2)_L \times SU(2)_R$
models which predict $t$-$b$-$\tau$ YU \cite{big-422}, where $t$ denotes the top quark. The low energy implications of $b$-$\tau$   \cite{Chattopadhyay:2001mj} and $t$-$b$-$\tau$  \cite{bigger-422} YU have been discussed
in the recent literature. For instance, $t$-$b$-$\tau$ YU is not realized in the mSUGRA/constrained minimal supersymmetric standard model (CMSSM) \cite{Chamseddine:1982jx}
because of the difficulty of implementing radiative electroweak symmetry breaking (REWSB) \cite{Baer:2009ie}.

In this paper we explore the low energy consequences of implementing $b$-$\tau$ YU in the CMSSM framework. REWSB in this case is not an issue anymore.
We refer to this combination of $b$-$\tau$ YU and CMSSM as YCMSSM, the `Yukawa' constrained version of CMSSM. Among other things, we require that YCMSSM
delivers a viable cold dark matter (DM) candidate (lightest stable neutralino) whose relic energy density is compatible with the WMAP measured value \cite{Komatsu:2008hk}.
One of our main observations is that the allowed fundamental parameter space of CMSSM is strikingly reduced in the YCMSSM setup.
We find that $M_{1/2} \ll m_0$, where $M_{1/2}$ and $m_0$ denote
 universal gaugino and scalar soft SUSY breaking masses respectively. Furthermore,
$b$-$\tau$ YU at the level of 10\% or better yields the constraint  $5\,{\rm  TeV} \lesssim m_0 \lesssim 20\,{\rm TeV}$. The supersymmetric
threshold corrections including finite loop corrections to the $b$ quark mass play an essential role here \cite{Rattazzi:1995gk}.

The lightest supersymmetric  particle (LSP) neutralino is essentially a bino, the spin 1/2 supersymmetric partner of the $U(1)_{Y}$ gauge boson, which is closely followed
in mass by a slightly heavier (next to lightest sparticle for short NLSP) stop, a scalar partner of the top quark. The desired LSP relic abundance
is achieved via  neutralino stop  co-annihilation, which in our case requires that the NLSP stop is about 10-20\% heavier than the neutralino.
The parameter $\tan\beta$, the ratio of the up and down Higgs VEVs, turns out to lie in a narrow range $35\lesssim \tan \beta \lesssim 40$.
The universal trilinear scalar coupling ($A_0$) is found to satisfy $|A_0 / m_0| \sim 2.3$.

We highlight some LHC testable benchmark points with comparable LSP neutralino  and light stop masses of around $100-330 \, {\rm GeV}$, while the corresponding
 chargino and second neutralino masses are $200-600\, {\rm GeV}$ and gluino mass $ \sim0.6-1.7 \, {\rm TeV}$.
Together with the lightest SM-like Higgs with mass  $114-124 \, {\rm GeV}$, these are the only `light' (LHC accessible) particles
predicted in this NLSP stop scenario with $b$-$\tau$ Yukawa unification and neutralino DM. The squarks of the first two families,
the heavy stop, the two sbottom particles,and the charged sleptons, all have large (multi-TeV) masses.

The fundamental parameters of CMSSM are:
\begin{align}
m_0,M_{1/2},\tan\beta,A_0 , \rm sgn(\mu)  \label{params}
\end{align}
where    $\rm sgn(\mu)$ is the sign of supersymmetric bilinear Higgs parameter.
All mass parameters are specified at $M_{\rm G}$.

We use the ISAJET~7.80 package~\cite{ISAJET} to perform random scans over
the CMSSM parameter space. ISAJET employs two-loop  renormalization
group equations (RGEs) and defines $M_{\rm G}$ to be the scale at which $g_1=g_2$. This is more than adequate as a few percent deviation
from the exact unification condition $g_3=g_1=g_2$ can be assigned to unknown
GUT scale threshold corrections~\cite{Hisano:1992jj}. The random scans cover the following parameter range:
\begin{align}
&0\leq  m_{0} \leq 25\, \rm{TeV}, &
0\leq M_{1/2}  \leq 2\,\rm {TeV}, \nonumber \\
&1.1\leq \tan\beta \leq 60, &
-3\leq A_{0}/m_0 \leq 3,
\label{parameterRange}
\end{align}
with $\mu >0$ and $m_t = 173.3\, {\rm GeV}$ \cite{:2009ec}. The results are not too
sensitive to one or two sigma variation in the value of $m_t$ \cite{Gogoladze:2011db}.

\begin{figure}[t]
\centering
\includegraphics[width=8.5cm]{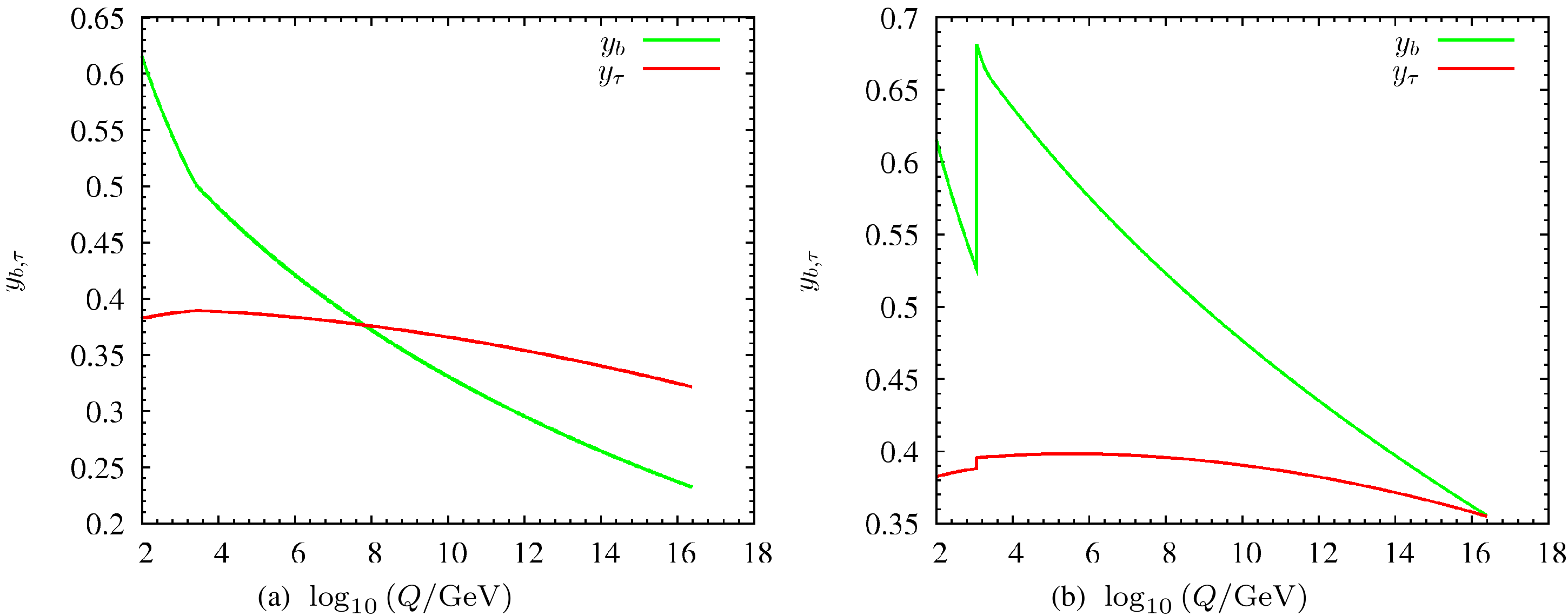}
\caption{
Evolution of bottom (green) and $\tau$ (red) Yukawa couplings without (a) and with (b) finite SUSY threshold corrections.
\label{plots1}}
\end{figure}
In scanning the parameter space, we employ the Metropolis-Hastings
algorithm as described in \cite{Belanger:2009ti}. All of the collected data points satisfy
the requirement of  REWSB,
with the neutralino in each case being the LSP. We direct the Metropolis-Hastings algorithm
to search for solutions with 10\% or better $b$-$\tau$ Yukawa unification (YU). After collecting the data, we impose
the experimental mass bounds on all particles~\cite{Amsler:2008zzb}, and use the
IsaTools package~\cite{Baer:2002fv} to implement the following phenomenological constraints:
\begin{table}[h]\centering
\begin{tabular}{rlc}
$m_h~{\rm (lightest~Higgs~mass)} $&$ \geq\, 114.4~{\rm GeV}$                    &  \cite{Schael:2006cr} \\
$BR(B_s \rightarrow \mu^+ \mu^-) $&$ <\, 5.8 \times 10^{-8}$                     &   \cite{:2007kv}      \\
$2.85 \times 10^{-4} \leq BR(b \rightarrow s \gamma) $&$ \leq\, 4.24 \times 10^{-4} \; (2\sigma)$ &   \cite{Barberio:2007cr}  \\
$0.15 \leq \frac{BR(B_u\rightarrow \tau \nu_{\tau})_{\rm MSSM}}{BR(B_u\rightarrow \tau \nu_{\tau})_{\rm SM}}$&$ \leq\, 2.41 \; (3\sigma)$ &   \cite{Barberio:2008fa}  \\
$\Omega_{\rm CDM}h^2 $&$ =\, 0.111^{+0.028}_{-0.037} \;(5\sigma)$               &  \cite{Komatsu:2008hk}
\end{tabular}
\end{table}

{\noindent}As far as  muon anomalous  magnetic moment   is concerned, we only require that the model does no worse than the standard model (SM).
In Fig.~\ref{plots1}(a) we show the evolution of
$b$ and $\tau$ Yukawa couplings  without  the SUSY threshold corrections
for a representative $b$-$\tau$ YU solution.
It is evident  that without suitable  the SUSY threshold corrections, $b$-$\tau$ YU occurs at around $10^{8}\,{\rm GeV}$.
In Fig.~\ref{plots1}(b)  we show  the need  for   SUSY threshold corrections in order to
achieve $b$-$\tau$  YU.

 The SUSY correction   $\delta m_{\tau}$  to the $\tau$  lepton
mass is given by $\delta m_{\tau}=v\cos\beta \delta y_{\tau}$.
 The finite and logarithmic corrections \cite{Pierce:1996zz} to the $\tau$ lepton mass are typically small,
{ so that the value of $y_{\tau}$ at $M_{\rm G}$ is more or less fixed.
From Fig.~\ref{plots1}(a), to implement  $y_{\tau}=y_{b}$ at $M_{\rm G}$ { therefore requires suitable
threshold correction ($\delta y_{b}$) to the bottom quark \cite{Rattazzi:1995gk}}.
The dominant contribution to $\delta y_b$ comes from
the gluino and chargino loops  \cite{Rattazzi:1995gk,Pierce:1996zz}. With our sign conventions ({ the sign of $\delta y_b$
is fixed as  $y_b$  evolves   from $M_{\rm G}$ to $M_{Z}$}), a useful approximate formula for the finite one loop
correction to $\delta y_b$ is given by

\begin{align}
\delta y_b^{\rm finite}\approx\frac{\mu}{4 \pi^2}\left (\frac{g_3^2}{3}\frac{ m_{\tilde g}
}{m_{1}^2}+ \frac{y_t^2}{8}\frac{A_t}{m_{2}^2}\right ) \tan\beta.
\label{finiteCorrectionsEq}
\end{align}
Here $g_3$ is the strong gauge  coupling, $m_{\tilde g}$ is
the gluino mass, $A_t$ is the stop trilinear coupling, and $m_{1} \approx ({m_{\tilde b_1} +m_{\tilde b_2}})/2$,
$m_{2} \approx ({m_{\tilde t_2} +{\mu}})/2$. ${\tilde b_1}$, ${\tilde b_2}$ denote
the two bottom squarks,  ${\tilde t_2}$ is the heavier stop,  and we assume that $m_{\tilde g}\ll{m_{\tilde b_1,\tilde b_2}}$
and $m_{\tilde {t_1}}\ll \mu, {m_{\tilde t_2}}$}.

\begin{figure}[b]
\centering
\includegraphics[width=8.5cm]{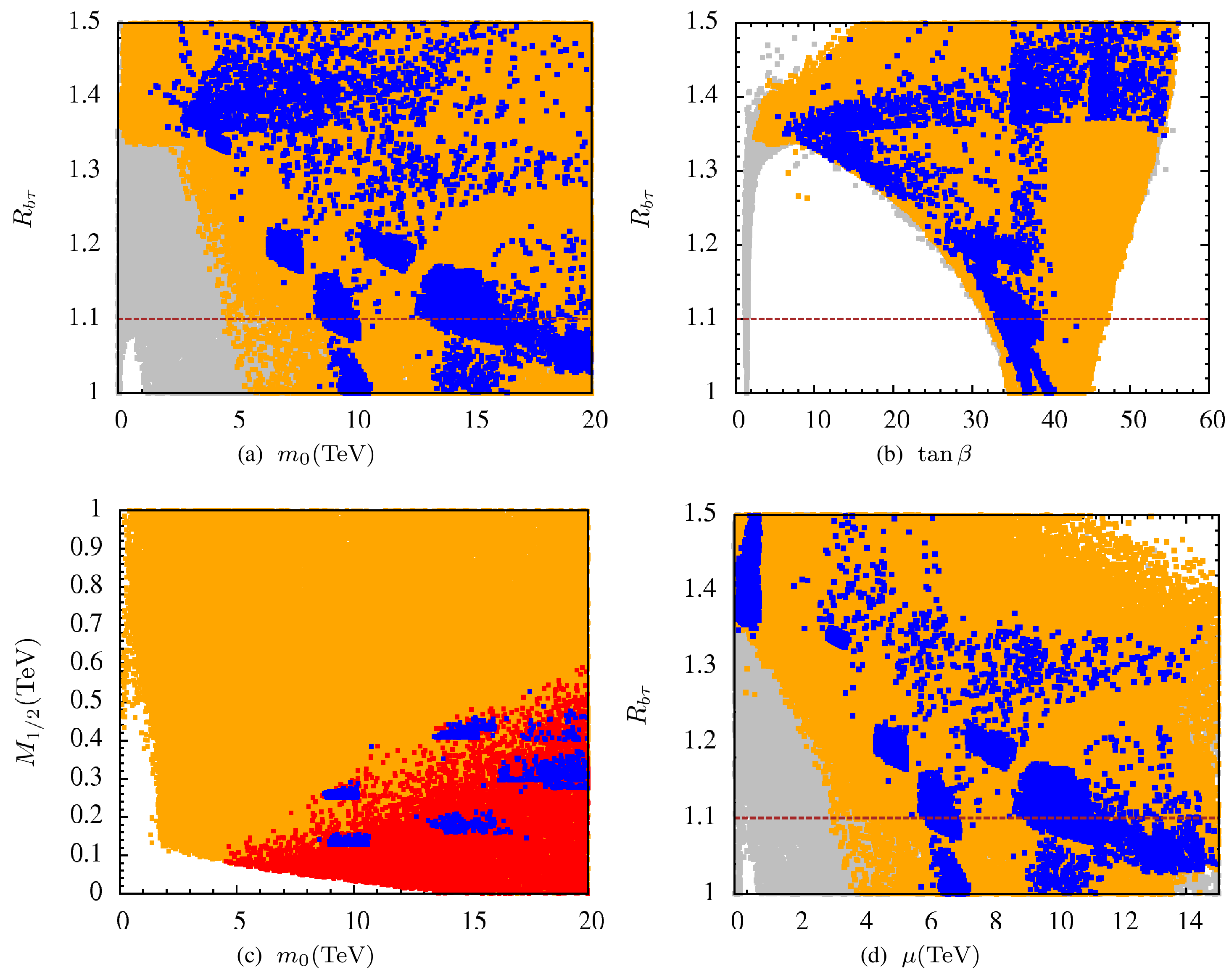}
\caption{
Plots in $R_{b\tau}$ - $m_0$, $R_{b\tau}$ - $\tan\beta$, $M_{1/2}$ - $m_0$ and $R_{b\tau}$ - $\mu$ planes.
In panels  (a,  b, d)   the gray points are consistent with REWSB and $\tilde{\chi}^0_{1}$  LSP. The orange points satisfy in addition  the particle mass
bounds, constraints from $BR(B_s\rightarrow \mu^+ \mu^-)$,  $BR(B_u\rightarrow \tau \nu_{\tau})$  and $BR(b\rightarrow s \gamma)$.
The blue points form a subset of orange points that satisfies WMAP bounds on $\tilde{\chi}^0_1$ DM abundance.
The dashed (red) line represents 10\% or better $b$-$\tau$ YU.
In $M_{1/2}$ - $m_0$ plane the orange color has the same meaning as described above, red color represents the solution
with 10\% or better $b$-$\tau$ YU, and  blue points satisfy the WMAP bounds.
\label{plots2}}
\end{figure}

In order to achieve $b$-$\tau$ YU,  $y_b$ must receive a negative contribution ($-0.2 \lesssim \delta y_b/y_b \lesssim -0.07$)
 from threshold corrections.  This is a relatively narrow interval compared to the full range
$-0.2 \lesssim \delta y_b/y_b \lesssim 0.25$  in the data that we have collected. The logarithmic
corrections to $y_b$ are  in fact, positive, which  leaves the finite corrections to provide for the correct
$\delta y_b$. Since $\mu>0$, the gluino contribution is positive, and so the
contribution from the chargino loop not only has to cancel out the contributions
from the gluino loop and the logarithmic correction, it also must provide
the correct overall (negative) contribution to $\delta y_b$. This is
achieved only for suitable large $m_0$ values and large negative
$A_t$, for which the gluino contribution scales as $M_{1/2}/m_0^2$ while the
chargino contribution scales as $A_t/m_0^2$.
Note that large values of $m_0$ imply heavy slepton masses, and so $b$-$\tau$ YU does
not provide any significant SUSY contributions to the muon anomalous  magnetic moment.

In order to quantify $b$-$\tau$ YU, we define the quantity $R_{b\tau}$ as
\begin{align}
R_{b\tau}=\frac{ {\rm max}(y_b,y_{\tau})} { {\rm min} (y_b,y_{\tau})}.
\end{align}
In Fig.~\ref{plots2}
we present our results
in the  $R_{b\tau}$ - $m_0$, $R_{b\tau}$ - $\tan\beta$, $M_{1/2}$ - $m_0$ and $R_{b\tau}$ - $\mu$ planes.
In panels  (a, b, d)   the gray points are consistent with REWSB and $\tilde{\chi}^0_{1}$  LSP. The orange points satisfy the particle mass
bounds, constraints from $BR(B_s\rightarrow \mu^+ \mu^-)$, $BR(B_u\rightarrow \tau \nu_{\tau})$ and $BR(b\rightarrow s \gamma)$.
The blue points form a subset of orange points that satisfies WMAP bounds on $\tilde{\chi}^0_1$ DM abundance. The dashed (red)
line represents 10\% or better $b$-$\tau$ YU. In the $M_{1/2}$ - $m_0$ plane the orange color has the same meaning as described earlier,
red color represents solutions with 10\% or better $b$-$\tau$ YU, and the blue points satisfy in addition the WMAP bounds.
In the $R_{b\tau}$ - $m_0$ plane of Fig.~\ref{plots2}, we show that in order to have 10\% or better
$b$-$\tau$ YU consistent with the collider bounds, we require
$m_0\gtrsim 5\,{\rm TeV}$.  But if $b$-$\tau$ YU is to be compatible with neutralino DM relic density,
represented by the blue points in the figure, we can see that $m_0\gtrsim 8\,{\rm TeV}$. To better appreciate  this result we
consider the $M_{1/2}- m_0$ plane. As we will see later, the neutralino-stop co-annihilation channel is the only
solution we have found for neutralino DM compatible with  $b$-$\tau$ YU.
The lighter stop mass can be as light as $100 \, {\rm GeV}$ \cite{Amsler:2008zzb}, which means that the neutralino mass has to
be  of  comparable  order,  in order to implement the stop co-annihilation solution.
An ISAJET  two loops  analysis  yields $M_{1/2}\gtrsim 150 \,{\rm GeV}$.
We can see from Fig.~\ref{plots2}(c) that $M_{1/2}\gtrsim 150 \,{\rm GeV}$ is compatible with
 10\% or better YU (points in red) for $m_0 \gtrsim 8\,{\rm TeV}$.

   In the $R_{b\tau}$ - $\tan\beta$ plane (Fig.~\ref{plots2}(b)) we have two regions with 10\% or better $b$-$\tau$ YU.
  In one region  $\tan\beta\approx 1.1$,
but this is ruled out by the light CP-even higgs mass bound.  The second region with successful $b$-$\tau$  YU  occurs for  $34\lesssim\tan\beta \lesssim 45$. { Beyond this range we do not have REWSB if we require 10\% or better $b$-$\tau$ YU.
Imposing , in addition, the neutralino DM requirement (represented by blue points), the allowed region shrinks to  $35\lesssim\tan\beta \lesssim 40$.

The plot in $R_{b\tau}$ - $\mu$ plane shows that for 10\% or better YU, $\mu \gtrsim 3\,{\rm TeV}$ (orange points).
The need for large values of $\mu$ can be understood by analyzing the threshold effects associated with the $b$ quark. As previously mentioned,
the second term in Eq. (\ref{finiteCorrectionsEq}) has to be larger than the first one for successful  $b$-$\tau$ YU.
With $\tan\beta$ squeezed in a relatively narrow interval, and the magnitude of $A_t$ bounded to avoid color and charge breaking,
needs  up with a large $\mu$ term to obtain  the required  $\delta y_b$.  On the other hand,
a bino-Higgsino mixed DM scenario requires $m_{\tilde{\chi}^0_{1}}\approx\lvert \mu\rvert$. From Fig.~\ref{plots2}(c),  the neutralino cannot be much heavier than $600 \,{\rm GeV}$ or so for successful $b$-$\tau$ YU. But with
$\mu > 4\,{\rm TeV}$  that a bino-Higgsino mixed DM scenario is not realized here.

 Next let us consider the A-funnel scenario of CMSSM where one needs $m_{A} \approx 2 m_{\tilde\chi_1^0}$. For large $\tan\beta$,  $m_{A}^2=2b/\sin(2\beta)\approx b\tan\beta$,  where $b$ is the soft SUSY breaking Higgs bilinear term. We have $b\tan\beta\approx 4\,{\rm TeV}$ for good $b$-$\tau$
unification,  which implies that $m_A\gg 2 m_{\tilde{\chi}^0_{1}}$ and thus, the A-funnel scenario is not consistent with
10\% or better $b$-$\tau$ unification.

\begin{figure}[b]
\centering
\includegraphics[width=8.5cm]{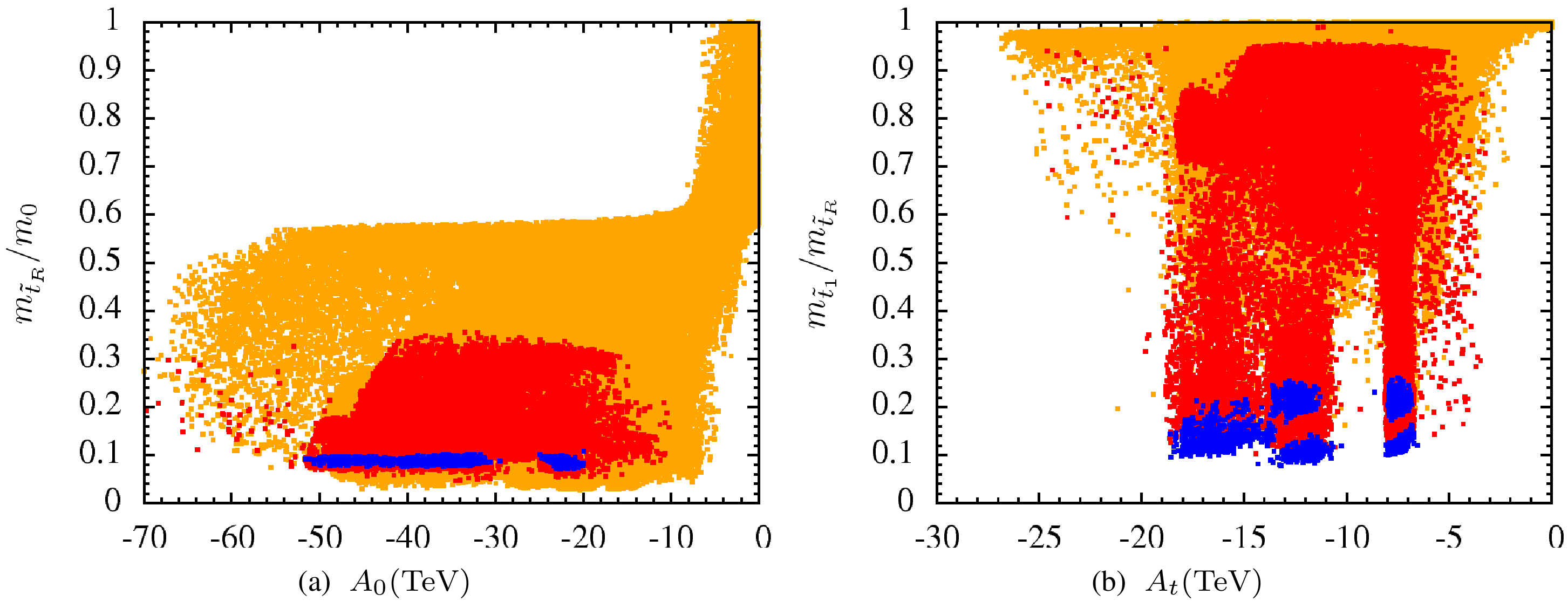}
\caption{
Plots in  $m_{\tilde t_R}/m_{\tilde U_R}-A_{0}$  and $m_{\tilde t_1}/m_{\tilde t_R}-A_{t}$
planes. The orange points satisfy the particle mass bounds, constraints from $BR(B_s\rightarrow \mu^+ \mu^-)$,
$BR(B_u\rightarrow \tau \nu_{\tau})$ and $BR(b\rightarrow s \gamma)$.  Red color represents the solution
with 10\% or better $b$-$\tau$ YU.
Blue color  represents the neutralino-stop co-annihilation solutions.
\label{plots3}}
\end{figure}

\begin{table}[h!]
\centering
\scalebox{0.8}{
\begin{tabular}{lccc}
\hline
\hline
                 & Point 1 & Point 2 & Point 3       \\
\hline
$m_{0}$          & 15220   & 10040  & 17920         \\
$M_{1/2}$        & 177     & 152    & 521      \\
$\tan\beta$      & 37      & 39     & 37       \\
$A_{0}/m_{0}$    & -2.36   & -2.32  & -2.33           \\
$sgn(\mu)$       &  +1   & +1   & +1       \\

\hline
$m_h$            & 115  & 120  & 115      \\
$m_H$            & 6036 & 4566 & 9752       \\
$m_{A}$          & 5997 & 4537 & 9688       \\
$m_{H^{\pm}}$    & 6037 & 4568 & 9753       \\

\hline
$m_{\tilde{\chi}^0_{1,2}}$
                 & 124,272  & 97,209     & 290,592        \\
$m_{\tilde{\chi}^0_{3,4}}$
                 & 10379,10379 & 6836,6836    & 12347,12347    \\

$m_{\tilde{\chi}^{\pm}_{1,2}}$
                 & 275,10406  & 211,6840  & 598,1239   \\
$m_{\tilde{g}}$  &  796   & 640     & 1680   \\

\hline $m_{ \tilde{u}_{L,R}}$
                 & 15170,15214  & 10000,10030  & 17892,17942    \\
$m_{\tilde{t}_{1,2}}$
                 & 153,5930  & 114,4076  & 328,7894     \\
\hline $m_{ \tilde{d}_{L,R}}$
                 & 15170,15222 & 10000,10036 & 17892,17951    \\
$m_{\tilde{b}_{1,2}}$
                 & 6060,8357 & 4152,5752  & 8097,11159    \\
\hline
$m_{\tilde{\nu}_{1}}$
                 &   15223     &  10041     &  17929        \\

$m_{\tilde{\nu}_{3}}$
                 &   12744     &  8453     &  15082        \\
\hline
$m_{ \tilde{e}_{L,R}}$
                & 15211,15208    & 10032,10032 & 17911,17909    \\
$m_{\tilde{\tau}_{1,2}}$
                & 9843,12771    & 6619,8474  & 11801,15130    \\
\hline

$\sigma_{SI}({\rm pb})$
                & 3.28$\times 10^{-12}$ & 5.85$\times 10^{-12}$ & 7.93$\times 10^{-13}$ \\

$\sigma_{SD}({\rm pb})$
                & 3.90$\times 10^{-12}$ & 2.39$\times 10^{-11}$ & 1.77$\times 10^{-12}$ \\

$\Omega_{CDM}h^2$
                & 0.11      & 0.09    &  0.1     \\

$R_{b\tau}$
               & 1.00      & 1.02    &  1.09     \\
\hline
\hline
\end{tabular}
}
\caption{ Point 1 has perfect  $b$-$\tau$ unification  with $R_{b\tau}=1$,
point 2 represents a solution with relatively light stop mass ($\sim 114\,{\rm GeV}$),  and
 point 3 represents a solution with a heavier stop mass ($\sim 330\,{\rm GeV}$).
The remaining squarks and sleptons all have masses in the multi-${\rm TeV}$ range.}
\label{table1}
\end{table}

The neutralino-stop co-annihilation channel is compatible with 10\% or better $b$-$\tau$ unification,
as shown by the blue points in Fig.~\ref{plots2}. Let see how  a light stop mass ($m_{\tilde t_1}$)
of $O(100)\,{\rm GeV}$,  is realized for  $m_0\gtrsim 8\, {\rm TeV}$ and $M_{1/2}\lesssim 600\,{\rm GeV}$.
In this region of the parameter space the diagonal entries of the stop mass matrix  to pick up dominant contribution from $y_t$ and
$A_t$  couplings, thus making the $m_{\tilde t_R}$ entry lighter than its value at $M_{\rm G}$. In Fig.~\ref{plots3}(a) we see that
the ratio $m_{\tilde{t_R}}/m_0$ can be as small as 0.05 for large values of $A_0$, which means that at the SUSY scale $m_{\tilde t_R}$
can be as light as $1\,{\rm TeV}$ or so, despite the large $\sim (8-20)\,{\rm TeV}$ $m_0$ values. On the other hand,  from Fig.~\ref{plots3}(b) we see that at SUSY scale, the
value of $\lvert A_t\rvert$ can be  $O(7-20)\,{\rm TeV}$. The off-diagonal entries $-m_t(A_t+\mu \cot\beta)$ for the stop quark mass matrix can therefore
be of comparable magnitude to the diagonal entries. Because of this, as seen in Fig.~\ref{plots3}(b), one
of the eigenvalues ($m_{\tilde t_1}$) of the stop quark mass matrix can be 10-20 times smaller than
$m_{\tilde t_R}$ at the SUSY scale. Thus,   with $m_{\tilde t_1}$  O(100) ${\rm GeV}$, we can have neutralino-stop co-annihilation compatible with
$b$-$\tau$ YU. A similar discussion for the stau mass matrix shows that the stau co-annihilation scenario  is not realized in YCMSSM.

 In Table~\ref{table1} we present three characteristic benchmark points which satisfy all, especially the dark matter,  constraints.
Point 1 displays essentially perfect $b$-$\tau$ YU solution with $R_{b\tau}=1$, point 2 represents a solution with a relatively
light stop mass ($\sim 114\,{\rm GeV}$), and finally point 3 represents a solution with a  heavier stop mass ($\sim 330\,{\rm GeV}$).
We note that the remaining squarks and sleptons all have masses in the multi-${\rm TeV}$ range.

 Since  the LSP is essentially a pure bino, both its spin-independent and spin-dependent cross sections on nucleons are
rather  small \cite{Gogoladze:2010ch},  $\sim 10^{-47}-10^{-48} \rm{ cm}^2$.
 Consequently it wont be easy  to detect the LSP in direct and indirect experiments.

In summary, we have investigated $b$-$\tau$ Yukawa unification in the mSUGRA/CMSSM framework and find that it is consistent with the
NLSP stop scenario and  yields the desired LSP neutralino relic abundance. This YCMSSM
predicts that there are just two `light' (LHC accessible) colored sparticles, namely the NLSP stop with mass $\sim 100-330\,{\rm GeV}$,
and the gluino which is $\sim 600-1700 \,{\rm GeV}$.  The chargino and a second neutralino are about a factor 2-3 lighter than the gluino.
The remaining squarks as well as all sleptons  have masses in the multi-TeV range. Regarding the fundamental CMSSM parameters,
we find that  $5\,{\rm TeV}\lesssim m_0 \lesssim 20\,{\rm TeV}$, $m_0/M_{1/2}\approx 30-50$,   $\tan\beta \approx 35-40$,
$\lvert \mu\rvert \sim 3-15 \,{\rm TeV}$ and $\lvert A_0/m_0\rvert \sim 2.2-2.4$.

We thank Rizwan Khalid, Azar Mustafayev and Mansoor Ur Rehman for valuable discussions. This work
is supported in part by the DOE Grant No. DE-FG02-91ER40626
(I.G., S.R. and Q.S.) and GNSF Grant No. 07\_462\_4-270 (I.G.).

\end{document}